\documentclass[twoside]{article}
\usepackage[latin1]{inputenc}
\usepackage{amssymb}
\usepackage{graphicx}

\usepackage[german,english]{babel}
\usepackage{multicol}
\usepackage{vmargin}
\setpapersize{A4}
\setmargnohf{35mm}{45mm}{132mm}{220mm}

\begin{document}

\title{A Machine-Independent Port of the SR Language Run Time System to
NetBSD Operating System} 

\author{Ignatios Souvatzis\\
	University of Bonn, CS Dept., Chair V\\ {\tt
	<ignatios@cs.uni-bonn.de>}}
\date{29th of September, 2004}
\maketitle
\thispagestyle{empty}

\section{Introduction}

SR (synchronizing resources)\cite{srbook} is a PASCAL -- style language
enhanced with constructs for concurrent programming developed at the
University of Arizona in the late 1980s\cite{srold}.
MPD (presented in Gregory
Andrews' book about Foundations of Multithreaded, Parallel, and
Distributed Programming\cite{MPDbook}) is its successor, providing the
same language primitives with a different syntax.

The run-time system (in theory, identical) of both languages provides
the illusion of a multiprocessor machine on a single single- or multi-
CPU Unix-like system or a (local area) network of Unix-like machines. 

Chair V of the Computer Science Department of the University of Bonn
is operating a laboratory for a practical course in parallel programming
consisting of computing nodes running NetBSD/arm, normally used via PVM,
MPI etc.

We are considering to offer SR and MPD for this, too. As the original
language distributions are only targeted at a few commercial Unix
systems, some porting effort is needed, outlined in the SR porting
guide\cite{sr:portingguide}.

The integrated POSIX threads support of NetBSD-2.0 should allow us to use
library primitives provided for NetBSD's phtread system to implement 
the primitives needed by the SR run-time system, thus implementing 13
target CPUs at once and automatically making use of SMP on VAX, Alpha,
PowerPC, Sparc, 32-bit Intel and 64 bit AMD CPUs.

This paper describes work in progress.

\section{Generic Porting Problems}

Given the age of the software and the gradual development of the C
language and the operating system environments available, some
adaptation is to be expected. Fortunately, the latest distribution of
SR  (version 2.3.2) has already been portend to two relatively modern
Unix-like environments (Solaris 2.2 and Linux), so the necessary
changes turned out to be confined to a one area:


{\tt gcc 3}, the system compiler of NetBSD-2.0, doesn't provide old
{\tt <varags.h>} variable argument functions anymore, so those had to
be converted to {\tt <stdarg.h>} syntax. Also, none of those functions
had fixed arguments. Most of the functions had a first logical parameter
{\tt char *locn} which could be changed into a fixed parameter. A few
functions had a first integer parameter (a count of the remaining 
parameters). In one case ({\tt sr\_cat}), the loops extracting the 
parameters from the variable argument list had to be changed to be 
initialized with the newly introduced fixed parameter.

\section{Verification methods}

SR itself provides a basic and an extended verification suite for the
whole system; also a small basic test for the context switching primitives.

The basic suite should be run to test an installation; the context
switch tests and the extended suite are used to verify a new porting
effort.\cite{sr:portingguide}

\subsection{Context Switch Primitives}

The context switch primitives can be independently tested by running
{\tt make} in the subdirectory {\tt csw/} of the distribution; this
builds and runs the {\tt cstest} program, which implements a small
multithreaded program and checks for detection of stack overflows, 
stack underflows, correct context switching etc.

\subsection{Overall System}

When the context switch primitives seem to work individually, they --- 
and the building system used to build SR, and the {\tt sr} compiler, 
linker, etc. need also to be tested.

A basic verification suite is in the {\tt vsuite/} subdirectory of 
the distribution; it can be extended with more tests from a seperate
source archive {\tt vs.tar.Z}. It is run by calling the driver script
{\tt srv/srv}, which provides normal and verbose modes, as well as using
the installed vs. the freshly compiled SR system. The only test that is
expected to fail is the {\tt srgrind} source code formatter --- it needs
the {\tt vgrind} program as a backend.

\section{Performance evaluation}

SR comes with two performance ealuation packages. The first, for the
context switching primitives, is in the {\tt csw/} subdirectory of 
the source distribution; after {\tt make csloop} you can start
{\tt ./csloop N} where N is the number of seconds the test will run
approximately.

Tests of the language primitives used for multithreading are in the
{\tt vsuite/timings/} subdirectory of the source tree enhanced with 
the verification suite. They are run by three shell scripts to 
compile them, run them, and summarize the results in a table.

\section{Establishing a baseline}

There are two extremes possible when implementing the context switch
primitives needed for SR: implementing each CPU manually in assembler code
(what the SR project does normally) and using the SVR4-style
{\tt getcontext()} and {\tt setcontext()} functions which operate 
on {\tt struct ucontext}; these are provided as experimental code
in the file {\tt csw/svr4.c} of the SR distribution.

\begin{table}
\begin{center}
\label{cswtimes}
\begin{tabular}{|l|l|}
\hline Implementation & Context switch times\\
\hline
i386 assembler & 0.059 $\mu$s\\
SVR4 system calls & 6.025 $\mu$s\\
\hline
\end{tabular}
\end{center}
\caption{\em Raw context switch times}
\end{table}

The first tests were done by using the provided i386 assembler
context switch routines. After verifying correctness and noting 
the times (see tables \ref{cswtimes} and \ref{hltimes}), the same
was done using the SVR4 module instead of the assembler module.

All tests were done on a 500 MHz Pentium III machine with 16+16 kB of
primary cache and 512 kB of secondary cache, and 128 MB of main memory, 
running NetBSD-2.0\_BETA as of end of June 2004.

\begin{table}
\label{hltimes}
\begin{center}

\begin{tabular}{|c|r|r|}
\hline Test description & i386 ASM & SVR4 s.c.\\
\hline
loop control overhead			& 0.01 $\mu$s& 0.01 $\mu$s\\
local call, optimised			& 0.07 $\mu$s& 0.07 $\mu$s\\
interresource call, no new process 	& 1.45 $\mu$s& 1.39 $\mu$s\\
interresource call, new process 	& 2.95 $\mu$s& 22.20 $\mu$s\\
process create/destroy 			& 2.46 $\mu$s& 26.14 $\mu$s\\
\hline
semaphore P only 			& 0.07 $\mu$s& 0.07 $\mu$s\\
semaphore V only 			& 0.05 $\mu$s& 0.05 $\mu$s\\
semaphore pair 				& 0.11 $\mu$s& 0.11 $\mu$s\\
semaphore requiring context switch 	& 0.39 $\mu$s& 9.09 $\mu$s\\
\hline
asynchronous send/receive 		& 1.71 $\mu$s& 1.63 $\mu$s\\
message passing requiring context switch & 1.90 $\mu$s& 14.50 $\mu$s\\
\hline
rendezvous 				& 2.65 $\mu$s& 27.05 $\mu$s\\
\hline
\end{tabular}
\end{center}

\caption{\em Run time system performance. The median times reported by
the SR script {\tt vsuite/timings/report.sh} are reported.}
\end{table}

The table shows a factor-of-about-ten performance hit for the operations
that require context switches; note, however, that the absolute values
for all such operations are still smaller than $30\,\mu{}s$ on 500\,MHz 
machine and will likely not be noticable if a parallelized program is 
run on a LAN-coupled cluster: on the switched LAN connected to the test
machine, the time for an ICMP echo request to return is about 250 $\mu{}s$.

\section{Possible improvements using NetBSD library calls}

While using the system calls {\tt getcontext} and {\tt setcontext}, as
the {\tt svr4} module does, should not unduly penalize an application
distributed across a LAN, it might be noticable with local applications.

However, we should be able to do better than the {\tt svr4} module without
writing our own assembler modules, as NetBSD 2.0 (and up) contains its
own set of them for the benefit of its native Posix threads library 
({\tt libpthread}), which does lots of context switches within a kernel
provided light weight process (\cite{nathan:sa}). The primitives
provided to {\tt libpthread} by its machine dependent part are the two
functions {\tt \_getcontext\_u} and {\tt \_setcontext\_u} with similar
signatures to {\tt getcontext} and {\tt setcontext}.

There are a few difficulties that arise while pursuing this.

First, on one architecture (i386) {\tt \_setcontext\_u} and {\tt
\_getcontext\_u} are implemented by calling through a function pointer
which is initialized depending on the FPU / CPU extension mode available
on the particular CPU used (on i386, 8087-mode vs. XMM). from this. On
this architecture, {\tt \_setcontext\_u} and {\tt \_getcontext\_u} are
defined as macros in a private header file not installed. The developer
in charge of the code has indicated that he might implement public
wrappers; until then, we'd have to check all available NetBSD
architectures and copy the relevant files. 

Second, there is no context initializing function at the same level as
{\tt \_setcontext\_u} and {\tt \_getcontext\_u}. {\tt makecontext} looks 
like it would be good enough but this has to be analyzed further.

\section{Work items left to do}

\subsection{Building a package for {\tt pkgsrc}}

To ease installation, a prototype package for the NetBSD package system
has been built. It needs a bit of refinement, though, but will be
available soon. (As the NetBSD package system is available for more
operating systems than NetBSD, a bit more work is needed.)

\subsection{Implementing and testing multithreaded SR}

SR can be compiled in a mode where it will make use of multiple threads
provided by the underlying OS, so that it can use more than one CPU of a
single machine. This has not been implemented yet for NetBSD, but should
be.

\end{document}